\documentclass[prd,amsmath,notitlepage,twocolumn]{revtex4-2}
\usepackage{graphicx}
\usepackage{float}
\usepackage{epstopdf,cancel}
\usepackage{epsf,latexsym,bbm,euscript}
\usepackage{amssymb,amsmath}
\usepackage{mathtools} 
\usepackage{times,graphics}
\usepackage{soul,xcolor}
\usepackage{mathtools}
\usepackage{mathrsfs}

\usepackage[caption=false]{subfig}
\usepackage{scalerel}
\usepackage{tikz}
\usetikzlibrary{svg.path}
\definecolor{orcidlogocol}{HTML}{A6CE39}
\tikzset{
	orcidlogo/.pic={
		\fill[orcidlogocol] svg{M256,128c0,70.7-57.3,128-128,128C57.3,256,0,198.7,0,128C0,57.3,57.3,0,128,0C198.7,0,256,57.3,256,128z};
		\fill[white] svg{M86.3,186.2H70.9V79.1h15.4v48.4V186.2z}
		svg{M108.9,79.1h41.6c39.6,0,57,28.3,57,53.6c0,27.5-21.5,53.6-56.8,53.6h-41.8V79.1z M124.3,172.4h24.5c34.9,0,42.9-26.5,42.9-39.7c0-21.5-13.7-39.7-43.7-39.7h-23.7V172.4z}
		svg{M88.7,56.8c0,5.5-4.5,10.1-10.1,10.1c-5.6,0-10.1-4.6-10.1-10.1c0-5.6,4.5-10.1,10.1-10.1C84.2,46.7,88.7,51.3,88.7,56.8z};
	}
}
\newcommand\orcidlink[1]{\href{https://orcid.org/#1}{\mbox{\scalerel*{
				\begin{tikzpicture}[yscale=-1,transform shape]
				\pic{orcidlogo};
				\end{tikzpicture}}{X}}}}

\def\sg{\textsl{g}}
\def\cO{\mathcal{O}}

\newcommand{\be}{\begin{equation}}
\newcommand{\ee}{\end{equation}}
\newcommand{\ba}{\begin{eqnarray}}
\newcommand{\ea}{\end{eqnarray}}
\def\half{\tfrac{1}{2}}

\def\6{{\langle}}
\def\9{{\rangle}}
\newcommand{\pad}{\partial}
\def\etal{\textit{et al.}}

\newcommand{\defeq}{\vcentcolon=}
\newcommand{\eqdef}{=\vcentcolon}

\def\eF{\EuScript{F}}

\def\eN{\EuScript{N}}

\def\ren{\mathrm{ren}}

\usepackage{url,hyperref}
\hypersetup{colorlinks,linkcolor={blue!55!black},citecolor={red!50!black},urlcolor={blue!45!black},breaklinks=true}

\begin{document}
	
\title{Structure and statistical properties of the semiclassical Einstein equations}

\author{Daniel R.\ Terno}
\email{daniel.terno@mq.edu.au}
\affiliation{School of Mathematical and Physical Sciences, Macquarie University, NSW 2109, Australia}

\begin{abstract}
We treat the semiclassical Einstein equation as a quantum-classical hybrid and demonstrate the formal equivalence of its two derivation methods. This approach identifies the left-hand side of the equation as the expectation value of the Einstein tensor given the state of matter, and not its actual value in each realization of the set-up. As a result, standard criticisms of semiclassical gravity do not apply, and stochastic gravity emerges as a necessary extension
\end{abstract}
	
\maketitle

\section{Introduction}

The Einstein equations of general relativity
\be G_{\mu\nu}=8\pi T_{\mu\nu}, \label{claE} \ee
are possibly the most perfect expression of classical physics. As summarised in the famous aphorism of John Wheeler, space  (represented by the Einstein tensor $G_{\mu\nu}=R_{\mu\nu}-\half R\sg_{\mu\nu}$, with $R_{\mu\nu}$ and $R$ being the Ricci tensor and the scalar, respectively), tells matter how to move, and the matter in the form of the energy-momentum tensor (EMT) $T_{\mu\nu}$ tells space how to curve.

Mathematically, general relativity is the simplest member of a broader family of metric theories of gravity \cite{BLLD:18,DT:10}. Both theoretical and observational considerations indicate that general relativity is a  low-energy limit of an effective quantum gravity theory \cite{BLLD:18,DT:10,K:12,HV:20}. Despite this, Eq.~\eqref{claE} remains the fundamental tool for exploring gravitational phenomena, ranging from the post-Newtonian corrections of satellite trajectories to astrophysical and cosmological studies.

Predictions of all proposed quantum gravity theories are expressed in classical terms \cite{K:12,W:22}. The observable Universe is modeled using classical geometry, which forms the foundation for both the standard cosmological model and discussions of its tensions or alternatives \cite{AZ:22,AZ:23,W:11,P:00}.

As the rest of physics falls within the scope of quantum theory, astrophysics and cosmology ``routinely" \cite{W:22} combine quantum mechanical descriptions of matter and classical gravity. Quantum mechanics, whether in its non-relativistic form or as quantum field theory and particle physics, determines basic parameters --- typically expectation values --- that characterize matter and fields. When many-body properties of bulk matter become significant, statistical mechanics provides additional methods. These calculations are generally performed in flat spacetime, as the relevant scales are much smaller than the curvature scale. Then, some algorithm that expresses the equivalence principle incorporates flat spacetime results into equations valid in general relativity \cite{P:00,W:22}.

From a foundational viewpoint, the absence of experimental evidence for gravitational field quantization (see \cite{MV:24,TMBP:24} for discussions on experimental attempts) makes hybrid quantum-classical schemes plausible. These schemes necessitate an interface that defines a mathematically coherent relationship between functions describing geometry (e.g., $G_{\mu\nu}$) and operators characterizing quantum matter \cite{BT:88,PT:01,T:23}. A four-level hierarchy of models  represents coupling of quantum matter to classical gravity \cite{H:14}. It begins with the Newton–Schrödinger equation, which describes non-relativistic particles in weak gravitational potentials (level 0), progresses to quantum fields propagating on curved backgrounds (level 1), and includes semiclassical gravity (level 2). Beyond these, it encompasses stochastic semiclassical gravity, effective field theory approaches to matter-gravity systems, and models incorporating a minimal length scale expected from canonical quantum gravity or modified commutation relations.

Arguments by M{\o}ller \cite{M:62} and Rosenfeld \cite{R:63} suggest using the (renormalized) expectation value of the EMT operator as the source of the Einstein equations, leading to a mean-field quantum hybrid approach (level 2 in the above classification) \cite{H:14,HV:20}. This proposal results in the semiclassical Einstein equation (SCE),
\be
G_{\mu\nu} = 8\pi\6\psi|\hat T_{\mu\nu}|\psi\9_\ren, \label{SCE}
\ee
accompanied by the formal evolution equation,
\be
i\hbar \frac{\partial}{\partial t}\psi = \hat H[\hat\phi,\hat\pi,\sg]\psi, \label{schro}
\ee
where the Hamiltonian depends on all matter fields, their conjugate momenta, and the metric $\sg_{\mu\nu}$ (The Einstein tensor $G_{\mu\nu}$ is derived from this metric). On a given background, renormalization introduces finite terms quadratic in curvature. The coefficients of these terms must be determined, and their inclusion raises the system's order to the fourth. While solving such a system conceptually requires a self-consistent approach, practical calculations are often perturbative or rely on specific geometric backgrounds. In practice it is usually done as a perturbative treatment  on a chosen backgrounds \cite{HH:81,FW:96,PT:09}. Alternatively, emphasis may be placed on specific properties of geometry that enable the extraction of structural features of the solutions \cite{MMT:22}.

Before addressing the implications of particular experiments for semiclassical gravity, two significant foundational issues require attention. First, mixtures of matter states—and consequently mixtures of geometries—introduce additional complexities. Averaging geometric quantities across different spacetimes leads to gauge noncovariant results, which are only meaningful when there is a clear algorithm for resolving gauge freedom in all possible scenarios \cite{FW:96}.

Second, hybrid schemes for quantum-classical dynamics can broadly be categorized as reversible (unitary) and irreversible \cite{T:23}. The former aim to provide a mathematically consistent quantum-classical counterpart to fully quantum unitary and fully classical Hamiltonian theories, without introducing dissipation or diffusion. However, all known reversible schemes are in general inconsistent \cite{T:23}, and there is substantial reason to suspect this is not merely due to a lack of ingenuity \cite{GGS:23}. Moreover, one key insight from the studies of such models is that quantum matter complies with the Heisenberg uncertainty relations only if classical quantities are defined with a certain inherent uncertainty \cite{BCGG:12,T:23} .

 A useful perspective on hybrid schemes views them as resulting from applying a classical limit to only one subsystem within the combined quantum--classical system. From this viewpoint, the full quantum description encompasses the entire system, with the hybrid formally derived by introducing two Planck constants, $\hbar_\mathrm{c}$ and $\hbar_\mathrm{q}$. Setting $\hbar_\mathrm{q} = \hbar$ and taking the limit $\hbar_\mathrm{c} \to 0$ yields the hybrid dynamics \cite{CS:99}.

The SCE belong to this class. In the discussion below, we first demonstrate that the two formal methods for deriving Eq.\eqref{SCE} are equivalent when viewed as derivations of a hybrid equation. This derivation further indicates that, strictly speaking, only a statistical interpretation of the Einstein tensor as $\langle G_{\mu\nu} \rangle_\psi$ is viable. Subsequently, the generalization of Eq.\eqref{SCE} becomes straightforward. In this generalized form, the SCE are more challenging to falsify: they are automatically consistent with the results of the Page--Geilker experiment \cite{PG:81} and make it impossible to distinguish between proper and improper mixtures—a property recently discussed in the context of mean-field theories \cite{FK:24}.

In the following we set $c=1$, occasionally keep $G\neq 1$ and explicitly write the Planck constant. 

\section{Derivation of the quantum-classical hybrid}

Despite the reasonable form of Eq.~\eqref{SCE}, its derivation is nontrivial, and its interpretation remains somewhat contentious \cite{FW:96,K:12,HV:20}. This can be compared with the Newton--Schr\"{o}dinger equation, where the inclusion of the self-gravity term is highly intuitive. However, the Newton--Schr\"{o}dinger equation is not a one-particle weak-field non-relativistic limit of the SCE but rather an equation governing an effective mean-field wave function in the limit of an infinite number of particles \cite{AH:14}.

There are at least two distinct methods by which Eq.~\eqref{SCE} can be formally derived. To simplify the discussion, we consider the Einstein--Hilbert action for gravity and describe the matter content as scalar fields, possibly conformally coupled \cite{FW:96}.

The first method expands a quantum metric and quantum scalar field formally around a classical vacuum solution, deriving the equation of motion for the expected metric by retaining only the tree-level diagrams for gravitons and both the tree-level and one-loop diagrams for the scalar field. Discussion of the two counterterms, whose coefficients must be determined \cite{FW:96,F:05,PT:09}, as well as the physical conditions under which this approach is valid \cite{K:12,HV:20}, is beyond our current scope. However, it is important to note that the loop expansion in quantum field theory is effectively an expansion in powers of $\hbar$: tree diagrams contribute terms of order $\hbar^0$, while one-loop diagrams contribute terms of order $\hbar$.

The second method uses a formal device of considering \( N \) non-interacting scalar fields in the same quantum state and performing the loop expansions. The limit \( N \to \infty \) is then taken under the constraint \( GN = \text{const} \). In this expansion, graviton loops are suppressed relative to the matter loops. In the \( N \to \infty \) limit, fluctuations in the expectation value of the matter EMT become negligible, and Eq.~\eqref{SCE} emerges as the limiting equation.

We adopt the analysis of \cite{HH:81}, which provided the original realization of this approach. Consider the transition amplitude between two states specified by a particular three-geometry, represented as a three-metric $^3\sg$ in a specified gauge, and the values of $N$ scalar field configurations $\vec{\phi} = \{\phi_j\}_{j=1}^N$ on the initial and final hypersurfaces $\Sigma'$ and $\Sigma''$, respectively. This amplitude is expressed as the path integral
\be
\big\langle ^3\sg'',\vec{\phi}''|^3\sg',\vec{\phi}'\big\rangle = \int_{\Sigma'}^{\Sigma''} \! \!\breve{D}[\sg] D\big[\vec{\phi}\big] \exp\left(\frac{i}{\hbar}\big(S_\sg[\sg] + S_m[\sg,\vec{\phi}]\big)\right). \label{pathA}
\ee
Here, the fields are constrained to take prescribed values on the initial and final surfaces. The measure includes the four terms $\eF_\alpha$ that represent the gauge conditions, along with the Faddeev-Popov determinant $\Delta_\eF$:
\be
\breve{D}[\sg] = {D}[\sg] \prod_{\alpha=0}^4 \delta\big(\eF_\alpha[\sg]\big)\Delta_\eF[\sg],
\ee
while the infinite gauge volume factor has been omitted.

The gravitational action includes \cite{F:05} the Einstein--Hilbert term, cosmological constant, two counterterms and the boundary term that we are not writing out explicitly,
\begin{align}
S_\sg&=\frac{1}{16\pi G_0}\int\!d^4x\sqrt{-\sg} \left(R-2\Lambda_0+\alpha_0 R^2+\beta_0 R_{\mu\nu}R^{\mu\nu}\right) \nonumber\\
&+\mathrm{boundary~terms},
\end{align}
where $G_0$ and $\Lambda_0$ are the bare values of the gravitational constant and the cosmological constant, respectively, and $\alpha_0$ and $\beta_0$ are additional bare coupling constant. The divergent parts of $\6\hat{T}_{\mu\nu}\9_\psi$ are removed by redefinition of these constants. Their finite renormalized values are physical parameters of the theory \cite{F:05,FW:96}.

The $N$ identical massless conformably coupled real scalar fields,
\be
S_m=-\frac{1}{2}\sum_{j=1}^N\int\!d^4x\sqrt{-\sg}\left(\pad_\mu\phi_j\pad^\mu\phi_j+\tfrac{1}{6}R\phi_j^2\right),
\ee
but it is immaterial for our argument. The exposition is further simplified if one assumes that the states of scalar fields are described by the same initial and final configurations, $|\phi'\9$ and $|\phi''\9$, respectively.

The evaluation of the path integral \eqref{pathA} proceeds in two step. First, for a given metric  and for each field configuration the functional $Y_{\phi'',\phi'}[g]$ via
\be
\exp(iY_{\phi'',\phi'}[g])\defeq\6\phi''|\phi'\9_\sg=\int\!D[\phi]e^{iS_m[g,\phi]/\hbar}.
\ee
Then the full amplitude becomes
\be
\big\langle ^3\sg'',\vec{\phi}''|^3\sg',\vec{\phi}'\big\rangle= \int_{\Sigma'}^{\Sigma''}\!\!\! \breve{D}[\sg]  \exp\left(\frac{i}{\hbar}\big(S_\sg[\sg]+NY_{\phi'',\phi'}[g]\big)\right).
\ee
Taking the $N\to\infty$ limit requires rescaling of the gravitational coupling such as
\be
NG\eqdef\kappa=\mathrm{const}, \label{scale}
\ee
 that brings the amplitude to the form
\be
\big\langle ^3\sg'',\vec{\phi}''|^3\sg',\vec{\phi}'\big\rangle= \int_{\Sigma'}^{\Sigma''}\! \! \breve D[\sg]\exp\left(\frac{iN}{\hbar}\big(S_\sg[\sg]+Y_{\phi'',\phi'}[g]\big)\right).
\ee
For large $N$ the
dominant contribution to the functional integral comes from metrics near
the extremum of  $\Gamma_{\phi'',\phi'}[\sg]\defeq S_\sg[\sg]+Y_{\phi'',\phi'}[\sg]$. Here $S_\sg$ is the classical action, and $Y_{\phi'',\phi'}[\sg]$ is (analogous to the) effective action, that for non-interacting scalar field is precisely given by the Gaussian expression. For interacting field one-loop expressions provide the leading corrections.

The connection to $1/N$ power counting is particularly transparent in the toy model of Ref.~\cite{HV:20}, whose action is
\begin{align}
S_\sg&+S_m= -\frac{1}{G}\int\!d^4 x\left(\pad_\mu h\pad^\mu h+h\pad_\mu h\pad^\mu+\ldots\right) \nonumber \\
&-\half\sum_{j=1}^N\left(\pad_\mu\phi_j\pad^\mu\phi_j+\ldots\right)+\sum_{j=1}^N\left(h\pad_\mu\phi_j\pad^\mu\phi_j+\ldots\right).
\end{align}
The self coupling graviton term of the order  $\cO(h^3)$ which appears in perturbative gravity beyond the linear approximation is explicably included.

 The computation of the dressed graviton propagator includes several types of diagrams. The limit \( N \to \infty \) is again needs to be accompanied by the rescaling of Eq.~\eqref{scale}. The first Feynman diagram is the free graviton propagator, which is now of order $\cO(\kappa/N)$. Next, there are $N$ identical diagrams with one loop of matter and two graviton propagators as external legs. Presence of two graviton propagators consignes them to the order $\cO(\kappa^2/N^2)$, hence the overall contribution can be represented by a single diagram of order $\cO(\kappa^2/N)$. The combined effect of all diagrams with two loops of matter and three graviton propagators is of the order $\cO(k^3/N)$, and so on.

The contributions that involve graviton loops are even more suppressed. A diagram with one graviton loop and two graviton legs contains four graviton propagators and two vertices. As the propagators contribute factors $(\kappa/N)^4$ and the vertices $(N/\kappa)^2$, this diagram is of the order $\cO(\kappa^3/N^2)$. As a result, in the limit $N \to \infty$, there are no contributions from the graviton propagators while the matter fields are quantized and contribute accordingly.

The scaling relations of the toy model as well as of the both more rigorous derivations of the semiclassical equation become nearly automatic if we keep $N=1$ (or just keep any finite number of various matter field), but proceed as appropriate in the derivation of the hybrid dynamics. We formally introduce two Planck constants (for gravity and matter fields, respectively) and take the classical limit for the gravitational sector only. It is accomplished by setting $\hbar=\eN\hbar_\sg=\mathrm{const}$, hence
\begin{align}
\exp\left(i(S_\sg +S_m )/\hbar\right)&\to\exp\left(i(S_\sg/\hbar_\sg +S_m /\hbar)\right)\nonumber \\
&=\exp\left(i(\eN S_\sg+S_m)/\hbar\right),
\end{align}
with $\eN\to\infty$ realizing the partial classical limit. Then suppression of the non-classical gravitational contributions becomes obvious.

\section{Statistical interpretation}
The SCE  is obtained by calculating expectation values of various correlation functions and then taking the limit $\hbar_\sg \to 0$. Hence, the meaning of the left-hand side of the SCE is:
\be
\6G_{\mu\nu}\9_\psi = 8\pi \6\hat{T}_{\mu\nu}\9_\psi^\ren. \label{SCE1}
\ee
This interpretation was posited by Ballentine in his commentary on the Page--Geilker experiment \cite{B:82}. Here, we see that this is a direct consequence of the way the SCE is derived.

It provides a resolution to the question of how the post-measurement state update rules \cite{BLPY:16} affect the SCE. It is convenient to discuss this, as well as the interpretation of the Page--Geilker experiment, in terms of the toy model proposed by Unruh \cite{U:84,K:12}.

The experiment can be understood as an elaboration of Schrödinger's cat gedankenexperiment. Two macroscopically distinct mass configurations are determined by the state of a decaying unstable particle. Denoting the state where the particle has not yet decayed, with the mass in one configuration, as $|0\9 \equiv |n\9|L\9$, and the state of a decayed particle with the other mass configuration as $|1\9 \equiv |d\9|L\9$, the quantum state after the approximate exponential decay of the unstable particle evolves as:
\be
\psi(t) = \alpha(t)|0\9 + \beta(t)|1\9,
\ee
where
\be
|\alpha|^2 \approx e^{-\lambda t}, \qquad |\beta|^2 \approx 1 - e^{-\lambda t}.
\ee
A Cavendish torsion balance is employed to measure the gravitational field induced by the mass configurations.

Initially $\6\hat{T}_{\mu\nu}\9(0)=\6 0|\hat{T}_{\mu\nu}|0\9$. Assuming that the exchange term is negligible, which is the case for macroscopically distinct configurations, we have
\be
\6\hat{T}_{\mu\nu}\9(t)\approx e^{-\lambda t}\6 0|\hat{T}_{\mu\nu}|0\9+(1-e^{-\lambda t}\6 1|\hat{T}_{\mu\nu}|1\9.
\ee
Hence, according to the SCE \eqref{SCE}, the Cavendish balance would follow the dynamics of the expectation value. In the experiment \cite{PG:81}, the balance suddenly moved to the position consistent with the distribution $L$ (when it entered the future light cone of the detection) after the counter clicked. Thus, Eq.~\eqref{SCE} is falsified as an adequate description of gravity \cite{PG:81,B:82,U:84}. Nevertheless, semiclassical gravity at this level does not provide any information about individual events. The correct SCE \eqref{SCE1} is still in agreement with the experiment.

In the light of Eq.~\eqref{SCE1}  the choice between ``no-collapse" (i.e. the many-world interpretation of quantum mechanics) and contradiction with the Bianchi identity \cite{PG:81,EH:77} does not arise. It is forced by the observation that in general if
\be
\6\psi|\hat{T}^\mu_{~\nu}|\psi\9_{;\mu}=0,
\ee
the measurement-induced discontinuity in  the state description will produce the corresponding discontinuity in $G^\mu_{~\nu\mu}$. However, this problem does not arise with Eq.~\eqref{SCE1}, and neither a possibility of a superluminal communication.

This clarification  does not matter much in actual applications. The SCE is usually used in astrophysical and cosmological problems that do not involve quantum states is a
superposition  of substantially different EMT distributions \cite{W:22}.  Nevertheless, this statistical interpretation trivially matches the results of \cite{PG:81}. It also gives a clear indication why stochastic gravity is necessary if one wants to capture the effect of fluctuations up to the second order \cite{HV:20}.

If this is the case, then the nonlinear relations between curvature and metric make its precise identification from the knowledge of  $\6G_{\mu\nu}\9_\psi$ impossible. Similar ambiguities in the averaging procedure are source of trouble for cosmology \cite{W:11,AZ:23}. The equations below can describe both application of some averaging procedure to inhomogeneous geometry, or evaluating exsections for a random process that is generating geometries (under assumption of the appopriate gauge fixing that ensures compatibility of different situations). Averaging both sides of Eq.~\eqref{claE} results in
\be
\6G^\mu_{~\nu}\9=\6R^\mu_{~\nu}\9-\half\delta^\mu_\nu\6\sg^{\lambda\rho}R_{\lambda\rho}\9=8\pi\6T^\mu_{~\nu}\9
\ee
One can focus on the averaged metric introduces $\bar{\sg}_{\mu\nu}\defeq\6\sg_{\mu\nu}\9$, and introduce
\be
\delta \sg_{\mu\nu}\defeq \sg_{\mu\nu}-\bar\sg_{\mu\nu}.
\ee
Then it is possible to define the connection $\bar\Gamma^\lambda_{~\mu\nu}$ and other objects that are based on the averaged metric. Then the averageds Einstein equations can be written as
\be
\bar G_{\mu\nu}+\delta G_{\mu\nu}=8\pi\6 T_{\mu\nu}\9,
\ee
where non-zero value of the correction term indeicates that the Einstein tensor of the averaged metric is not  in general the average of the Einstein tensors, $G[\6\sg\9]\neq \6G[\sg]\9$.

Most general states in quantum theory are mixed states --- convex combinations (i.e., the weighted averages) of some pure states, that the extreme points of the set of all quantum states. In the SCE framework the geometric quantities are certain real-valued functionals of pure  quantum states. Hence the standard rules of quantum mechanics lead to the formal expression
\be
\6G_{\mu\nu}\9_\rho=8\pi\mathrm{tr}\left(\rho \hat{T}_{\mu\nu}\right)_\ren=8\pi \sum_i \6\psi|\hat T_{\mu\nu}|\psi\9_\ren,  \label{SCEf}
\ee
where the mixed state is
\be
\rho=\sum_i w_i|\psi_i\9\6\psi_i|, \qquad w_i\geqslant0.
\ee
As discussed above there is no direct relationship between $\6G_{\mu\nu}\9_\rho$ and $\bar \sg_{\mu\nu}=\6\sg_{\mu\nu}\9_\rho$. This is so even if the EMT of each of the mixture components has low variance, so $\6\delta\sg\9_{\psi_i}\approx 0$. However, in this case
\be
\6\sg_{\mu\nu}\9=\sum_i w_i\6\sg_{\mu\nu}\9_{\psi_i},
\ee
does not have to correspond to any of the geometries in the ensemble.

A formal expression for the averaged geometric quantities, such as Eq.~\eqref{SCEf}, can be given more operationally-meaningful form if the all quantities are describe by relative to a reference frame that is constructed according to a pre-defined algorithm using relational quantities.  Statistical moments of the invariant quantities, such as independent curvature scalars \cite{SKMHH:03}, can be calculated directly.

This interpretation of the SCE  satisfies the requirement of operational indistinguishability between proper and improper mixtures \cite{BLPY:16} that was analysed by Fedida and Kent \cite{FK:24}. The mass configuration can be decided either as a result of a random process (a proper mixture) or as tracing out of the auxiliary degrees of freedom (an improper mixture).  In terms of the Unruh model the matter density matrix is in both cases
\be
\rho=|\alpha|^2|L\9\6L|+|\beta|^2|R\9\6R|.
\ee

In the example of Page-Gailker experiment the proper mixture corresponds to the ensemble of the post-measurement configurations (when the results are not revealed), and the improper mixture results from tracing out the unstable particle.
Assume the Eq.~\eqref{SCE} holds. Once
 the Cavendish balance is in the future light cone of the random choice of the matter configuration it will respond in one of the two possible distinct ways. On the other hand, in case of the improper mixture, the balnce will behave as (time-dependent) weighted average of the two responses.

Such behaviour would allow to distinguish between two types of mixtures. However, as the only valid prediction of the SCE is the expectation value $\6 G_{\mu\nu}\9_\rho$ such  differentiation is impossible. In addition, in the Newtonian limit if we assume that each of the two matter configurations r results in a low dispersion value of $\6\hat{T}_{00}\9_{L,R}$ (i.e. the states are approximate eigenvalues of the Hamiltonian \cite{K:12,U:84}), we will have that the Newtonian gravitational potential satisfies
\be
\6\varphi\9_\rho\approx|\alpha|^2\varphi_L+(1-|\alpha|^2)\varphi_R,
\ee
where $\varphi_{L,R}$ are the gravitational potentials that correspond the respective mass distributions.

\section{Discussion}

We have shown that the two formal derivations of the SCE are equivalent when viewed as derivations of the quantum-classical hybrid with $\hbar_\sg \to 0$. A key consequence of this derivation is that the Einstein tensor is fundamentally a stochastic quantity, and the SCE predicts only its expectation value. Sharp (low-dispersion) predictions are possible only under special conditions. As a result, unlike Eq.~\eqref{SCE}, it does not produce any of the undesirable effects discussed in Refs. \cite{EH:77,PG:81,FK:24}.
This analysis also highlights the necessity of stochastic gravity \cite{HV:20}, even under the assumption that all gravitational field fluctuations originate from fluctuations in the quantum matter.

The SCE lacks matter-gravity entanglement and can, in principle, be falsified in experiments where the gravitational field is used to establish entanglement between two matter subsystems with negligible other interactions \cite{MV:24}. However, further investigation is required to determine how its low-energy implications affect the interpretation of tabletop experiments.

 \acknowledgments Useful discussion with David Ahn, Samuel Fedida, Masahiro Hotta, Adrian Kent and Sebastian Murk are gratefully acknowledged.

\end{document}